\documentclass[11pt]{article}
\usepackage{graphicx}
\usepackage{amsmath}
\usepackage{amssymb}
\usepackage{amsxtra}
\usepackage{changebar}
\usepackage{placeins}

\newcommand{\bc}{\begin{center}}
\newcommand{\ec}{\end{center}}
\newcommand{\bd}{\begin{displaymath}}
\newcommand{\ed}{\end{displaymath}}
\newcommand{\be}{\begin{equation}}
\newcommand{\ee}{\end{equation}}
\newcommand{\ba}{\begin{array}}
\newcommand{\ea}{\end{array}}
\newcommand{\bea}{\begin{eqnarray}}
\newcommand{\eea}{\end{eqnarray}}
\newcommand{\bt}{\begin{tabular}}
\newcommand{\et}{\end{tabular}}

\newcommand{\bp}{\begin{picture}}
\newcommand{\ep}{\end{picture}}
\newcommand{\bfi}{\begin{figure}}
\newcommand{\efi}{\end{figure}}
\begin{document}


\title{{\huge \bf Seeking a Game in
which the standard model Group shall Win
  }}

\author{
H.B.~Nielsen
\footnote{\large\, hbech@nbi.dk}
\\[5mm]
\itshape{
The Niels Bohr Institute, Copenhagen,
Denmark}\\
Don Bennett
\itshape{Brooks Institute}
}

\date{}

\maketitle

\begin{abstract}
It is attempted to construct a
group-dependent
quantity that could be used to single out
the
Standard Model {\em group}
$S(U(2) \times U(3))$ as being the
``winner'' by this quantity being the
biggest possible
for just the Standard Model group. The
suggested quantity is
first of all based on the inverse
quadratic Cassimir for the fundamental
or better smallest faithful representation
in a notation in which the adjoint
representation quadratic
Cassimir is normalized to unity. Then a
further corrrection
is added to help the wanted Standard
Model group to win and the
rule comes even to involve the abelian
group $U(1)$ to be
multiplied into the group to get this
correction be allowed.
The scheme is suggestively explained to
have some physical
interpretation(s). By some appropriate
proceedure for extending the group
dependent quantity to groups that are
not simple we find a way to make the
Standard Model Group the absolute
``winner''. Thus we provide an indication
for what could be the reason for the
Standard Model Group having been chosen
to be the realized one by Nature.
 \end{abstract}

\newpage
\thispagestyle{empty}
\section{Introduction}
It is one of the great questions asked
in connection with
our Bled Conference: Why Nature has
selected just those gauge
groups, which we find? Of course so far
the only gauge group
found is that of the Standard Model.
Thus it is a priori this
gauge group, which we should attempt to
explain; then the theory, we might
invent for that purpose, may or may not
suggest further gauge groups
as for instance the hierarchy of gauge
groups suggested in the model of
Norma Mankoc et al.\cite{Norma}.
One of us (H.B.N) and Rugh and Surlykke
\cite{Rugh} estimated quantitatively
the amount of information contained in the
knowledge of the gauge group, and with
N. Brene\cite{Brene} we found that
defining a quantitative concept
of skewness - lack of automorphisms -
appropriately we could delare the Standard
Model Group to be characterized as
essentially the most ``skew''.

In the
present article we should
 go for inventing a somewhat different
{\em group
dependent quantity} than the ``skewness''
\cite{Brene}, and then imagine
that Nature for some reason has selected
just that group, which,
say,{\em  maximizes} this group dependent
quantity. This means that we strictly
speaking in a phenomenological way
attempt to adjust the rules of
a competition between groups and seek to
adjust the rules, so as
to make an already selected winner, the
Standard Model Group, to  win the
game. It is a bit as a great dictator
seeking to make, say, his son
become the winner of a sport game by
cleverly adjusting the rules of the
game, so that he wins. In an analogously
``nepotisitic'' way we shall seek
to arrange the game so as to make the
Standard Model group win the game.

It should be said though, that in
inventing the game we also look at
some physical model behind, much inspired
indeed by our long ongoing
project of Random
Dynamics\cite{RD,RDrev,Foerster}.
Strictly speaking there may be even
a couple
of routes suggesting, what groups are
``best'' based on the ideas of
Random Dynamics, that the fundamental
laws of Nature are extremely complicated.
That is to say, if indeed the laws of
nature were fundametally in some way
random and very complicated, what would
then be the characteristic property
-the strength so to speak - of the group
comination that would appear as the gauge
group effectively as seen by relative to
say Planck scale physicists working at
low energy?:

a) The one route is based on,  that the
gauge symmetry appears at first
by accident {\em approximately}, but that
then quantum fluctuations
take over and cause the gauge group to
appear effectively as
an exact gauge group \cite{Foerster}.
In such a philosophy
the group with the best chance should be
the group for which
a gauge theory most easily can appear
just by accident. Suggestively
such a favoured group should be one for
which, say a lattice
gauge theory, could most easily turn out
to appear with approximately
gauge invariant action by accident
\cite{gaugeglas, Foerster}. This
in turn is at least suggestively
argued to occur for a gauge group, for
which the range in the configuration
space, over which the action has not to
vary according to the rule
of gauge symmetry is in some way -
that may be hard to make precise
though - so {\em small as possible}. If
namely the range over which
the variation of the action shall be
small is ``small'', then there is
the better chanse to get it constant
approximately there just by accident.
This argumentation is then turned into
saying, that the range of variation
of the link variables caused by a gauge
transformation say associated
with a site in the lattice should be as
``small'' as possible in order to
make the gauge group most likely to occur
by accident. Now we typically
imagine the lattice link variables to be
or at least be represented
as matrix elements of some representation
of the gauge group. Well, at least
we typically take the action contribution
from one plaquette to be
a trace of some representation of the
gauge group. Normally we have the
``intuitive'' or conventional
expectation that although the most general
action contribution $S_{\Box}$ should
be a linear expansion on
traces/characters for all the possible
representations of the gauge group,
the traces of the smallest
representations would somehow dominate
this  expansion.  Such an expected
dominance of the small representation
trace in the action means that the
variation of the action as function
of the physical combination of the link
variables - i.e. the plaquette
variables - vary relatively slowly over
the group. But if we can get
the action in this sense vary relativly
slowly over the gauge group,
it may mean that it also suggestively
varies relatively slowly, when we
vary the
gauge. If somehow we have a ``setting''
- meaning say that everything
is written basically in terms of matrices
in some low representation
- so that the variation of the action
along the group is relatively
slow, then very likely one would think
that also the variation
in this ``setting'' of any possible
candidate for a term as a function
of some gauge transformation would a
priori be relativly slow. In other
words we say that a good `setting'' for
making the variation a priori along
the gauge variation ``small'' is one in
which the plaquette action
is dominated by a ``low''
representation/character
 - meaning say low
quadratic Casimir for it -.

It should be had in mind that the
quadratic Casimir is crudely a measure
for how much variation the representation
matrix in the representation in
question varies as function of the group
element it represents. If then
we imagine that in the lattice model,
which supposed to be the fundamental
model,
the group is represented by a certain
representation rather than directly
as an abstract group element, the
variation of the ``fundamental'' lattice
model variables are in some sense - that
may not be so clear though -
more slowly varying as function of the
group elements the smaller the
quadratic Casimir for the representation
functioning as the fundamental
fields. But the slower this variation is
the less extensive region
is passed by the ``fundamental'' lattice
variables and the easier
it would therefore be that by accident
under a gauge transformation
the action that were at first not taken
to be gauge invariant
would be so by accident nevertheless. By
this argumentation it is
here argued in words that may really be
meaningless that a
{\em a small quadratic Casimir for the
representation which
is used by Nature as the ``fundamental''
lattice field degrees
of freedom makes it more likely for the
gauge group in question
to occur by accident in an a priori
random action theory}.
The point should then crudely be that we
should look among
all groups and seek which ones have the
representations
with smallest qudratic Casimir for
representations that must still
be faithful in order to at all represent
the group in question.
The smaller these faithful
representations that could be used
the better should be the chance for the
group to be the one realized in Nature.

b) The second route - which were the one,
we started working on -
involves several assumtions which we have
worked on before, but it may
become too much for this route being
trustable unless we somehow can
get the number of assumptions somewhat
reduced. Most importantly
we assume ``Multiple Point Principle'' \cite{MPP} on
which we have worked much
and which states that there are several
degenerate vacua. That is to say
the coupling constants get -
mysteriously ? - adjusted so as to make
the theory discussed just sit on a phase
transition, where several
phases meet. The next assumtion then is
that after such an adjustment
of the lattice action coefficients -
which are basically the coupling
constants being adjusted by the
``Multiple point principle'' - we look
for that group which gives us the
numerically biggest value in the
vacuum realized (we argue it is the
Coulomb phase one) of the plaquette
action $S(U(\Box))$. For this
latter assumption we may loosely
say, well it means minimizing the energy
density may be. Or we may
involve the complex action model
\cite{complex} and
argue that a big contribution
for the plaquette action may likely lead
to a big contribution also
numerically for the imaginary part of the
action. Since now it is the main
point of this complex action model to
minimize the imaginary part of the
action the best chance for a certain
gauge theory to be realized
should then be, if it can give the
numerically biggest imaginary part.
But assuming real and imaginary parts to
depend in roughly the similar
way on the variables this would then
favour groups with that numerically
large plaquette action. We shall go into
this a bit complicated route
to get a suggestive game for the groups
in section \ref{MPProute}.

The game proposed at the end in the
present article is a somewhat
new one, but actually one of the present
authors (H.B.N.) and Niels
Brene\cite{Brene} long ago had a slightly
different
proposal for the game
to be won by the Standard Model
{\em group}, namely that it should
be the {\em most skew} in certain sense,
which we even made quantitative.
Of course ``skew'' for a group means that
it has relatively few
automorphisms. Honestly speaking we did
not yet publish what to
do with the Abelian invariant subgroups
so we strictly seaking
took the competion between the goups
having just one $U(1)$ invariant
subgroup. The precise quatity for the
game to minmize were then
the number $\# Out(G) $ of outer
automorfisms divided
by the logarith of the rank $r$
of the group. In reality what comes to
count a lot in this game
about skewness turns out to be the
division out of subgroups
of the center, which is what distinguishes
the various groups
having the same Lie algebra. We shall see
below that in order to
finally adjust the game of the present
article based in stead
at first on the quadratic Casirmir for a
faithful and small
representation to really get the Standard
Model group win
is again to allow this
``division out of a subgroup of the
center''. This  means the distingtion of
the
{\em group} (rather than
the Lie algebra) is to give a lot of
points
in the game. So at the end
we might be forced to let the game depend
much on the property of the
group rather than of the Lie algebra, and
that may presumably be the main
lesson that it is {\em the group property
rather than the Lie algebra
properties, that really matters to
select the Standard Model}.

We shall therefore in the foloowing
section
\ref{group} review the
seemingly so important distingtion
between group and Lie algebra,
and call attention to that we even though
we can
claim that  the
phenomenology of Standard Models gives us
not only the Lie algebra but also the
Lie group, so that this distingtion
really has a phenomenological
significance in fact in terms of the
representations of quarks and
leptons. In the following section
\ref{scaling} we shall then discuss
that at least reasonable notation
independent quanties have to be
chosen for the competition, so that the
game will not vary unreasonably
by varying notations and normalizations.
This suggests essentially to
use the Dynkin index   which is precisely
being an important index,
because it is somewhat sensible with
respect to independence of
notation as a start. Then in the next
section
\ref{scores} we shall
present the group theoretical values of
interest for our proposed game,
i.e. the Dynkin indices \cite{Dynkinindex}essentially and
the corrections connected with
the group rather than the Lie algebra for
at first the simple groups.
How to combine the simple groups by a
kind of averaging may open
up for a bit freedom and therefore
nepotism to let the Standrd Model
Group win, but really there is not so
much to do to help the
Stand Model Group, it must essentially
fight for itself. This discussion
is put into section \ref{averaging}. The
conclusive discussion of the game
is put into section \ref{gameconclusion}.  The model behind of the
somewhat more complicated nature
involving ``multiple point principle''
is put in section \ref{MPProute}. A by
itself very interesting
motivation for our a bit complicated
multiple point principle route
is, that it goes in connection with very
old attempts of ours to fit
the fine structure constants.

\section{ Phenomenological
significance of Group rather
than Lie Algebra}
\label{group}
A priori one might say that it is only
the gauge Lie algebra
of the Yang Mills theory that matters,
since the
 Yang Mills field theories
are constructed alone from the knowledge
of the Lie algebra
of the gauge group. So from this point of
view one
can say that the
Standard Model group (without now
stressing the word group it means
that we think of the Lie algebra) is
$U(1) \times SU(2) \times SU(3)$.
However, we can, and we shall in this
article, asign a ``phenomenological
meaning'' to the gauge {\em group} rather
than just the Lie algebra
by associating  the choise of the
{\em group} (among the
several groups having the same Lie
{\em algebra}) with the system
of representations under  which the
various matter fields - the Fermions
and the Higgs fields -  transform. The
reader should have in mind
that while all the possible
representations
for quarks and leptons and the Higgs
or  thinkable additions
to the Standard Model are allowed
a priori, we may prevent some by
requiring representation of a certain
{\em group}.
Indeed it is only some of the
representations of the {\em Lie algebra}
of the Standard Model, as we might
denote  the Lie algebra
of $U(1)\times SU(2) \times SU(3)$,
which are also representations of the
various {\em groups} having the same
Lie algebra, such as $U(1)\times
SU(2)/Z_2 \times SU(3)/Z_3$, $S(U(2)\times
U(3))$, $U(2) \times SU(3)$ etc.
For example the group
$SO(3)=SU(2)/Z_2$ has the same Lie
algebra as
$SU(2)$, but as is rather wellknown
while $SU(2)$ has all the representations
of the Lie algebra
- it is indeed the {\em covering group}
of say $SO(3)$ -both with
half integer and integer (weak iso)spin,
the group $SO(3)= SU(2)/Z_2$
has {\em only} as true representations the
(weak iso)spin integer
ones. Since the left handed quarks and
leptons belong to the
weak isospin =1/2 representation of
$SU(2)$, which is not allowed
as true representation of $SO(3)$, we can
 conclude that a group
with the same Lie algebra as the Standard
Model using $SO(3)$ instead
of $SU(2)$ would be an example of a
{\em group} that could {\em not}
be used in the Standard Model. It would
e.g. {\em not} be allowed
to claim that $U(1) \times SO(3)\times
SU(3)$ were the Standard Model
{\em group}, because it could not have
the left handed quarks and leptons
and the Higgs as representations.

So you see that there are many groups
that are forbidden as
Standard Model groups, but e.g. the
covering group ${\bf R} \times
SU(2) \times SU(3)$ for which all
representations of the Lie algebra
are also allowed representations of the
(covering) group could
at first not be prevented as
``the group for the Standard Model''.

However, it is our philosophy to impose a
{\bf phenomeological extra requirement
to select the group}, which deserves to
be called the {\em Standard
Model Group} (SMG). The idea is to among
the various groups with the
Standard Model Lie algebra, which are
allowed in the sense of having
all the representations present in the
Standard Model, we believe in,
to select as the Standard Model Group
to be that one (or several ?)
which is {\em most informative} w.r.t.
selecting, which representations
are allowed, so that just knowing this
group tells us as much as
possible about, which representations
occur in nature as presently known.
With requirement of the {\em most
informative}
group about the representations
in the Standard Model we should of course
not accept the covering
group ${\bf R}\times SU(2) \times SU3)$,
which would give no information,
provided we can at all find a group with
the Standard Model Lie algebra
which would exclude some representations
(which of course should be
some representations not found in nature
so far). Such a more informative
group giving correct information about
representations found empirically
is the group denoted
$S(U(2) \times U(3))$ =
$(U(1)\times SU(2) \times SU(3))/Z_6$. The
 symbol $U(2)$ in  this symbol
$S(U(2) \times U(3))$
alludes to it being constructed as a pair
of a $2\times2$ unitary matrix
(meaning one in the group $U(2)$) and
the $U(3)$ symbol alludes to an
$3\times 3$ unitary matrix (i.e.
one in $U(3)$) and then  the extra
condition
being imposed by  the $S$
in front that the product means
that  the
determinant of the two unitary
matrices put into a
$5\times 5$ matrix shall be unity. Seen
in this way it is rather
obvious that the here proposed
``Standard Model \underline{Group}''
$S(U(2)\times U(3))$ is a subgroup of
$SU(5)$ as
a group and not only as far as the Lie
algebra is
concerned. One can even say that some of
victories
of $SU(5)$ concerning the weak
hypercharges of the
particles in the Standard Model can be
ascribed to
the information gotten out of the from
$SU(5)$
surviving sub{\em group}
$S(U(2)\times U(3))$.
The second way of denoting the same
{\em group} $S(U(2)\times U(3))$ is
$U(1)\times SU(2) \times SU(3)/Z_6$
and it
describes it as
first considering the {\em group}
$U(1)\times SU(2) \times SU(3)$
and then divide out its center a certain
subgroup isomorphic to the
group of integers counted modulo 6, called
here $Z_6$. This special
subgroup is generated by the group element
$(2\pi, -{\bf 1}, \exp(2\pi/3) {\bf 1})$
of $U(1)\times SU(2) \times SU(3)$
and the elements generated by it being
divided out. This means that one divides
out the invariant subgroup generated
by this element $(2\pi, -{\bf 1}, \exp(
2\pi/3))$ so as to construct the
corresponding factorgroup.
 We here counted the length around
of the $U(1)$ as being $6*2\pi$, so that
the sixth power of the
generating element
$(2\pi, -{\bf 1}, \exp(2\pi/3) {\bf 1})$
becomes
the unit element in  $U(1)\times SU(2)
\times SU(3)$.
One might also describe this group
starting from the covering group
${\bf R}\times SU(2) \times SU(3)$
dividing out the subgroup generated
by essentially the  same element
 as we just used $(2\pi, -{\bf 1},
\exp(2\pi/3){\bf 1})$.

It should be remarked that by this
division out of
group isomorphic to the integers modulo
6 we get the three
invariant Lie {\em algebras} for
respectivly $U(1)$,
$SU(2)$, and $SU(3)$ linked together.
While the Lie group
$U(1) \times SU(2) \times SU(3)$ is the
cross product
of three factors, the suggested
phenomenological {\em group}
for the Standard Model, or for nature we
could almost say,
$S(U(2)\times U(3)) =
(U(1)\times SU(2) \times SU(3))/Z_6$
is {\em not} a cross product of any
corresponding groups.
This corresponds to that the rules for
hypercharge quantisation
which follows from the
``phenomelogically''supported group
$S(U(2) \times U(3))$ are such that the
hypercharge values $y/2$
allowed by this {\em group} depends on the
representations of the
non-abelian Lie algebras $SU(2)$ and
$SU(3)$.

It should be remarked immediately that
this type of bringing
an abelian group $U(1)$ together with
non-abelian groups
by division out of a discrete subgroup is
a rather characteristical property
of the Standard Model Group $S(U(2)\times
U(3))$. That means then that it is
``tallent for
the Standard Model Group'', in the sense
that among Lie groups
with similar rank or similar dimension as
this Standard
Model Group'' there are not many that
can  claim to divide
out in the nontrivial way a bigger
discrete group than this
$Z_6$, which is divided out in the
Standard Model Group case.
So if we want to ``help'' the
``Standard Model Group''
$S(U(2)\times U(3))$ to win a game, we
should it give
many points to have such a
``divission out'' with a relatively
large group, so that $S(U(2)\times U(3))$
can win on its
$Z_6$.

For example in the article by one of us
and N. Brene and one of us \cite{Brene} in
which we claimed that having few
automorphisms was what singled out the
Standard Model Group $S(U(2)\times U(3))$
among other groups with the same number
of abelian dimensions, it were in reality
the divission out of the discrete subgoup
$Z_6$ causing a connection between
SU(3) and U(1) that removed some seperate
automorphism acting on U(1) seperately
and one on SU(3) seperately replacing
it by only a common automorphism for them
both that helped to make the Standard
Model Group more skew so as to win the game for being ``skewest''.

\section{Introductory guidance
for what game
to propose}
One could imagine several directions for
speculations giving ideas
about what type of games among groups one
should attempt in order
to seek a game suitable for the Standard
Model Group to
win.

Some such inspiration ways of thinking
could be:
\begin{itemize}
\item One idea would be that the Standard
Model Group
is the end or close to the end of a
perhaps long series
of group break downs - you could think of
Normas theory
in which it comes after several break
downs of some $SO(N, 1)$
at higher energies- and thus one could
almost in Darwinistic
terms think about what would be the
typical way for a group
to break and under such a breaking, is
there some property that
gets enhanced by the breaking. By this we
mean: Is there
some property - expressed by number say -
of the group surviving the
break down that will typically or always
be bigger than for the group
that broke down to it. If we have such a
quantity we would - if
it is true that there are many
breakings - expect it to be so big for
the Standard Model Group that making a
game for such a quantity
would likely  make the Stadard Model win
or at least get close to win.
There are of course some quantites that
do get say smaller each time
the group breaks, namely the dimension or
the rank. So in such a
many breaking philosophy we would expect
that the Standard Model
Group would have - in some sense - very
low dimension and very low
rank, say. But it is difficult to say
what to compare.
At least we must admit that some groups
have smaller rank and/or
smaller dimesnion than the Standard Model
Group, so these simple ideas
were not quite so usefull.

One route though might be to require for
instance that the gauge group
we look for should have a system of Weyl
fermions that are both
mass protected and nevertheless leads to
no anomalies in the gauge
charges. Then one could even add (extra)
assumtions about
that the representions of the Weyl
fermions be in some sense small
or simple.

\item Alternatively we could think
somewhat in the direction of
the landscape model (from string theory)
\cite{landscape}
that there are many
a priori possible vacua having different
gauge groups. Then we need some
extra speculation or assumption about
which of these vacua then have
the best chanse of be the one in which we
come to live,
or which gets realized at all. To selects
such vacuum and
thus the gauge group to be found, one
might first think of the
antropic principle: then it would we
should speculate about
which gauge group would be the most
favourable for humans.

One could also say we need a theory for
inital conditions to tell
us which vacuum should be selected to be
produced in the beginning
and then likely survive. Here the complex
action model of
one of us(H.B.N.) and Ninomiya could come
in as a
candidate to select a vacuum. In fact the
main point of
this complex action modelends up being
that the initial
conditions get settled in such a way as
to minimize the
imaginary part of the action $S_I$
evaluated for the whole
history of the Universe though both past
and future. Since
so enormously much of the universe is
practically empty
- i.e. vacuum - it is clear minizing such
an imaginary
part of the action $S_I$ will in very
first approximation
mean that {\em that} vacuum should be
selected to exist through
most of time and space, which has the
smallest imaginary
part of the Lagrangian density
${\cal L}_I$. Without
knowing what the imaginary part of the
Lagrangian density
in the correct fundamental theory is it
is of course somewhat
difficult to guess how to get this
imaginary part of the
Lagrangian density minimized, but we
could attempt the
following loose argument: Suppose that
the imaginary part
of the Lagrangian density has a very
similar form as the
real part, just with different
coefficients. Then we would
guess that to find the minimal imaginary
Lagrangian density
we might instead seek an extremal real
part, and then hope
that this is where most likely the
extreme imaginary part
will also be. Such a search for an
extremal real
part of the Lagrange density might by
itself be supported by
other arguments without using complex
action model.
In fact we could speculate that somehow
the most stable
vacuum were one with smallest energy
density. Ignoring or
approximating away the kinetic part of
the energy density
extremizing the energy density would
lead to extremizing
the Lagragian density among the possible
vacua.

Such a search for a numerically largest
plaquette action -
if one thinks i a lattice gauge theory
model terms - could
thus be an idea that could be supported
by several
speculations; either our selection by
complex action model
or by some minmization of energy or
Lagrangian density.

But the Lagrangian density in a certain
vacuum of course
depends on the coupling constants or
equivalently on the
coefficients to the variuos terms that
may occur in
say a lattice gauge theory. Therefore
such a minimization
of the plaquette action among the
different vacua requires
that we have in addition a method for
calculating
these coefficients or coupling constants
for all the
different theories with their different
gauge groups,
which were what were to compete. Now at
this point
we propose our determination of the
coupling constant by
means of our principle of multiple point
principle
(MPP)\cite{MPP}. This principle MPP
means that the vacuum sits
at a phase transition point as function
of these coupling constants. But now it
sounds, that we have really
put too many unrelyable assumptions on
top of each other
so that the chanse of the all being
true gets very low:
existence of an imaginary action, vacuum
being selected
to have it minimal, the imaginary action
of the vacuum
being minimal just when real part is
extremal too, the multiple
point principle of couplings being chosen
to just sit
on the phase borders(at some multiple
point, where several
phases meet). And then to make use of
this long series of
assumptions we have to make the
approximations to be used
to estimate the size of the plaquette
action under the MPP
etc, assumpions. It is actually this
series of ideas
that were the point of the route of
section \ref{MPProute}.
But probably it can only be excused by
saying, that doing such
a series of speculations we have at least
an attempt to a connected picture and
should have a better chanse of
stumpling on to a correct proposal for
the game, that
is characterizing the Standard Model
Group, because
we should not make any totally stupid and
wrog things, if
we are in some at least thinkable scheme.

\item One very attracktive way to proceed
would be
a  genuine Random Dynamics\cite{RD,RDrev,
Foerster} . In principle
we might
imagine a quantum field theory, which
instead of being assumed translational
invariant is assumed to
have a quenched random (glassy) Lagragian
density\cite{gaugeglas} or action for the unit cell if
we think of the
model as regularized to let us say a
lattice type theory. We may even take
the number of degrees of freedom to vary
in a quenched random
way from cell to cell in the lattice. So
we take it, that there is connected to
each 4-cube
in a lattice at random - quenched
randomly -
chosen a number of degrees of freedom.
Next also in the quenched random way an
action contribution expression is chosen,
and that expression delivers then the
action contribution
from the cell considered, and it depends
of course from assumed locallity only on
the degrees of freedom of that cell and
the neighboring few cells. Such a model
on a lattice and with locallity and
background geometry put in but otherwise
with quenched random action and number
of degrees of freedom
 could be considered  a Random Dynamics
model\cite{RD,RDrev,gaugeglas} .

According to our old idea \cite{Foerster}
there can in  say a lattice model occur
effectively gauge invariance without it
being put in to the extend that a photon
without mass can appear in a model with
no exact gauge invariance. Let us though
mention that this
phenomenon of a gauge symmetry appearing
by itself, as one might say, comes about
that at a stage we write the theory as
having a formal
gauge symmetry looking at first as if it
were
Higgsed. Then it is the Higgs degrees of
freedom
in the formally gauge invariant
description,
that by quantum fluctuations wash out, so
as to become ordinary massive particles
or just
an unimportant field not accessible at
low energy. The idea should now be that
the quenched random theory proposed here
as a
manifestation of the Random Dynamics
project would in a way similar to the one
described in \cite{Foerster} be
rewrittable into a theory
with some formal gauge invariance, which
then due to quantum fluctuations could
appear at the end  not Higgsed (although
 it looked
at first Higgsed).
Thus some gauge symmetries would come out
as observable at long distances, giving
rise to say
massless photons gluons etc..

Thinking in terms of such a quenched
random theory producing effective
although at first
formal gauge symmetries it becomes in
principle a matter of a may be hard - but
presumably doable - computer calculation
to find out
which gauge groups occur and how offen
 in this
``by itself way''\cite{Foerster}, provided
though that we put in the definite rules
for the quenched random distribution of
the action and the number of degrees of
freedom per cell. However, it might very
likely turn out that this specific choice
of a quenched random distribution of the
degrees of freedom numbers per cell and
the action per cell will be of little
significance as to how the model will
show up at long
distances, and what guage groups will
appear.

Such an insensitivity to the details of
the quenched random probabilty set up
may though be just the wishful thinking
of Random Dynamics, that at the end it is
features of the theory determined by
looking
only at long distances (or in other
regimes,
where the ``poor'' physicist can get
access),
that determine the effective laws of
nature which we see. In any case it would
be a very
important project to by computer or just
theoretically find out which gauge groups
preferntially would come out by themselves
from such quenched random lattice theory
with even a quenched random number of
degrees of freedom (varying from cell to
cell).
In the spirit of the present article the
idea of course would be a bit
speculatively to figure
out what properties of a group would
make it likeliest that just that group
in question
would appear by itself.

How now to get an idea of which groups
would most
likely come out of such a quenched random
theory ?
Well, in order that one can get the
formally exact
gauge symmetry to appear effectively
so as to deliver massless gauge bosons
effectively it is needed, that
at the starting level the gauge symmetry
is there
approximately, because it is the rudiment
of the fundamentally not present gauge
symmetry being
broken, that leads to the ``Higgs'' or
Higgs like effects breaking at first
seemingly spontaneously the global gauge
symmetry, that has to be small enough
for being destructed by quantum
fluctuations. In other
words we only can get sufficient quantum
fluctuations
to bring the formal guage symmetry which
we might invent to become physically
effective at long
distances provided the original gauge
breaking
were small enough to be beaten by the
quantum
fluctuations. So we are in fact asking
for which gauge groups are likely to
occur by quenched
random accident in small regions of the
lattice theory
as {\em approximate } gauge symmetries.
Now let us think of seeking such a
locally accidentally
approximate gauge symmetry by starting
to look for it say near some starting
point in the configuration
space of the theory locally and then
estimate
the chanse, that going further and
further away from this point the action
will by accident not
change more than some limit corresponding
to the limit for getting it finally
appear as a long distance gauge symmetry.
Now a gauge transformation in a lattice
theory is to be thought about as
if we locally
have the possibility of transforming the
configuration
by means  of any  (gauge) group element.
So we now ask for how to get the best
chanse for that
we acting with any element in the group
corresponding to approximate realization
of the gauge symmetry
of the action at a certain site. When we
here talk
about a site, it is just meant that in
many places
one can presumably find some way of
transforming
the even random number locally of degrees
of freedom
in a neighborhood so as to approximately
(but
approximately only) not change the
action (contribution
from that region). To begin with in
asking for
approximate symmetry of the action at
first when the
gauge group elements of the
transformation are near
to the unit element it is mainly the Lie
algebra
that must be relevant. The chance for
having by accident the same action as
one goes further and
further away by transforming with
elements which lie
longer and longer away from the unit
element gets
of course smaller and smaller the longer
away we
go to ask for this accidental symmetry.
So it makes it most likely to find an
accidental symmetry for
a given group, when the action of the
group
changes the variables in the quenched
random theory as little as possible. In
the extreme case,
when the varibles of the quenched random
theory were not transformed at all the
invariance of the action would of course
be guaranteed,
but that would be a trivial case, that
would of course
at the end not lead to any effective
gauge theory at
long distances. So we must ask for a
slow variation,
but there should be some variation. To
make it easy
 - or at least for start -  we shall
think of the
degrees of freedom among the quenched
random ones
being roughly representation matrix
elements. That is
to say we may think of that there are
among the
quenched random number of degrees of
freedom locally
some we may think of as matrix elements
and of the
proposed transformation law as a linear
representation
of the group. In this way we allow
ourselves to think
of the speed with which the configuration
moves when
varying the group element in the (local)
gauge
transformation as motion speed for a
representaion
matrix. This latter speed is proportional
to the
square root of the quadratic
Casimir $c_R$ for the
representation in question $R$. So we see
that
the chance of getting an approximate
symmetry
under such a one point local gauge
transformation
is biggest, if the representation to which
we relate it
has the smallest quadratic Casimir,
because then
so to speak the speed of moving of the
configuration
- approximated by the matrix elements of
the representation $R$ - when we move
the group element
is the smallest. Since we thought of
starting
at around the unit group element and got
the
normalization for the speed to consider
specified
by the Lie algebra, we would naturally
count the
quadratic Casimir $c_R$ normalized by
setting the
quadratic Casimir for the adjoint
representaton,
which is the representation on the Lie
algebra itself,
equal to unity.

In this way we get from the Random
Dynamics picture of looking for
approximate gauge symmetry
by accident the suggestion of selecting
the game to be:

{\bf Which group has the smallest
quadratic Casimir for its smallest
faithful representaton
in a notation normalize to let the adjoint
represetation quadratic Casimir be
normalized to be $1$.
}

Typically of course it will be in the
local cases,
wherein the representation matrix that
we can
use as an approximation to the local
variables is one with the smallest
quadratic casimir that
will be most important for finding
approximate
gauge symmetry by accident, because it
is these cases
that have the biggest chanse. It is
therefore we in practice must think of
the relevant
representation $R$ as being the one with
the smallest
quadratic casimir. The representation
with smallest quadratic casimir is
practically the same
as the ``fundamental'' represntation.
Thus we
arrive essetially to that the game to win
for being the most likely group to appear
approximately by
accident is the one which has the
smallest fundamental
representation quadratic casimir $c_F$.
The ratio of
the two quadratic Casimirs, the
fundamental and the adjoint,
is actually such an interesting quantity
group theoretically
that it got essentially - i.e.
apart from some dimension of repressentation factors - the name
Dynkin-index.

In the spirit of the just above it is
clear that if
we could somehow ``divide out'' part of
the center
this would make the group smaller(in
volume) and
thus easier to
get realized as approximately a good
symmetry (i.e.
an approximate symmetry of the action) by
accident.
We should therefore let such a divission
out of the center
count extra, enhancing the success of the
group to win
the bigger the subgroup divided out.

As is explained a bit more in the
following section
\ref{scaling} it is suggested that we
should improve
our quatity to be minimzed to
$c_F/(\# \hbox{center divided out})^{2/d}$.
We can namely crudely consider $c_F$ as
proportional
to the $2/d$th power of the volume of the
group in the sense, that since $c_F$ is a
quadratic form in the ``distance'' in the
group the volume
of a d-dimensional group gets by varying
this
  $c_R$ from representation to
representation
its volume changed proportional to the
$d$th power
of the square root of $c_R$. If one
therefore
change the volume by some other effect
effefectively, namely by dividing out a
subgroup of the center
having $\# \hbox{centerpart divided out}$
elements - which
will of course diminish the volume by a
factor $\frac{1}{\# \hbox{center divided
out}}$, this
would correspond to replacing $c_F$ by
an effective quadratic Casimir $c_F/ \#
\hbox{centerpart divided out}$
\end{itemize}

\section{Requirements of correct
behavior under group
volume scalings}\label{scaling}
It is important to fix the precise
quantity to be proposed
as the one that the group winning should
say maximize so that this quantity shall
not be notation
dependent but as stable under change
of conventions as possible. It is
therefore we had to take the
ratio of two relatively easy to select
representations. If
we had namely
not taken a ratio this way the quadratic
Cassimirs would depend
on the notation for normalizing quadratic
Cassimirs.

For giving a possible good physical sense
to this ratio it is immediately
obvious that a meaning of the type that
this ratio denotes the square
of the speed of motion of the group
element in the two different
representations discussed is called for.
If now the true
physical quantity to be
argued for were indeed rather a total
volume ratio we can see that
a volume correction for say the
``fundamental'' representation
would have to come in  just the right
power to combine in a physically
consistent way with the speed ratio
already being present in the
proposed $1/c_F$. This considerations
leads rather quickly to
that our first proposal $1/c_f$ can only
be corrected by a division out of a
center subgroup of order $\# center$ by
the factor $(\# center)^{2/d}$, where $d$
is the dimension of the group.

That is to say that the quantity to be
say maximized would in order to combine
the volume dependence correctly
\begin{equation}
(\# center (divided out))^{2/d}/c_F.
\end{equation}
The to be minimized quantity could then
be of course the
inverse of this
$c_F/(\# \hbox{center divided out})^{2/d}$.

\section{What Scores do Different (Simple) Groups get?}
Before we in the next subsection
\ref{extraction} shall tell about how
one extracts from the litterature
the values for the quantity $1/c_F$,
we may put forward some features of
how the competition goes by mentioning
a few remarks:
\begin{itemize}
\item{{\bf Large rank behavior}}
As is wellknown the simple Lie algebras
are classified into four infinite
series and further some ``exceptional''
Lie algebras. For the infinite series
it actually turns out that if we allow
the smallest quadratic Casimir
representation $F$ to be the one making
$c_F$ smallest we get for the algebras
for ``large $N$'' - meaning the late
algebras in the infinite chain - that
\begin{equation}
\frac{1}{c_F} \rightarrow 2 \hbox{ for the
rank } r \rightarrow \infty.
\end{equation}

This is a very important property for
our project because you could add a
formally $\infty$ to the rank $r$ region
and the function of the algebra $1/c_F$
would remein a continuos function
and that now on a compactified space
of algebras. These means that there
should exist one (or perhaps several)
largest value for $1/c_F$. So we can
really expect to find a presumably single
winner among the simple algebras - or we
might have got an infinite limit, but
that luckily does not happen -.

\item{ {\bf  The front field}}
The winner number one among the simple
Lie algebras turrns out to be $SU(2)
= A_1$, since it gets using the general
formula for $A_1$
\begin{equation}
\frac{1}{c_F} = \frac{2}{1 -
\frac{1}{N^2}} \hbox{ for $SU(N)=A_{N-1}$,}
\end{equation}
that
\begin{equation}
\frac{1}{c_F(SU(2))} = \frac{2}{1 - 1/4} =
\frac{8}{3}.
\end{equation}
This 8/3 is the absolutely record for
any simple Lie algebra, and so
$SU(2) = A_1$ is the ``gold medal winner''
among simple Lie groups.

If we use the correctio factor
$(\frac{1}{\# \hbox{center-elements divided out}})^{\frac{2}{d}}$, which we mentioned
above it happens that it is also bigger
for $SU(2) = A_1$ than for any other
simple Lie algebra. In fact it is
for $SU(2)$ equal to $2^{2/3}$ =
1.587401052, so that the full score
with this factor included becomes for
the gold winner $SU(2)$ equal to
$ \frac{8}{3}*1.587401052 =4.233069472 $
So the winner just
even more certainly becomes $SU(2)$.

It is of course comforting for our
model that this absolute winner among the
simple Lie algebras is at least one of
the invariant subliealgebras of the
Standard Model Lie algebra $U(1) \times
SU(2) \times SU(3)$.

But now comes for our scheme a problem:
The silver winner among the simple Lie
algebras using only the ratio of the
quadratic Casimirs $\frac{1}{c_F}$ is
{\em not} as we might hope for the
Standard Model algebras to win the
$A_2 = SU(3)$ algebra, but rather
$SU(3)$ is beaten by $SO(5) = B_2
\approx Sp(4) = C_2$ which obtain
the score
\begin{equation}
\frac{1}{c_F(C_2)} = 12/5
\end{equation}
obtained from the general formula
$\frac{1}{c_F(B_r)} = \frac{4(r+1)}{2
r+1} = \frac{2N+4}{N+1}$ where
$N =2r$ by putting $r=2$ or
equivalently $N=4$. The $SO(5)=B_2$
Lie algebra is isomorphic to the
symplectic one $C_2$; to get the
fourdimensional representation,
which is the  vectorrepresentation
$V$ for the symplectic $C_2$, we
must for $SO(5) = B_2$ use the spinor
representation.

Now the for our hoped for explanation
of the Standard Model a bit unfortunate
fact is that the $SU(3) = A_2$ algebra
only reach the score $\frac{1}{c_F(A_2)}
= \frac{2}{1 - 1/3²} = \frac{9}{4} < \frac
{12}{5}$. So in the pure use of
$1/c_F$ the phenomenologically relevant
$SU(3) = A_2$ lost and only obtained
the bronce medal. Of course it is still
promissing that it got a medal at all,
but we could have said that we got
the two genuine simple Lie groups
if the winning gold and silver to be the
two phenomenologicallly found ones.
But alas, it were not like that
completely!

But now we have already mentioned the idea
of the extra factor $(\# center)^
{\frac{2}{d}}$, where $d$ is the dimension
of the algebra.

For $SU(3)$ and $SO(5)\approx Sp(4)$ the
extra factor turns out:
\begin{eqnarray}
\hbox{For $SU(3) = A_2$}& :&
(\# center)^{2/d} = 3^{1/4} = 1.316074013 \\
\hbox{For $Sp(4) = SO(5)=B_2 =C_2$}&:&
(\# center)^{2/d} = 2^{1/5} = 1.148698355.
\end{eqnarray}
Thus we get for the full scores when
this factor is included:
\begin{eqnarray}
\hbox{For $SU(3) =A_2$} & : &
\frac{(\# center)^{2/d}}{c_F} = 3^{1/4}*9/4
=1.316074013 *9/4 = 2.961166529 \\
\hbox{For $Sp(4)=C_2 = SO(5)=B_2$} &:&
  \frac{(\# center)^{2/d}}{c_F} = 2^{1/5}*
12/5 =1.148698355 *12/5 =2.756876052.
\end{eqnarray}
So we see that te extra factor from
dividing out the center just barely
brought the $SU(3)$ algebra in front
of $SO(5)\approx Sp(4)$ by .20 out of
ca 2.9 meaning by $7 \%$.

This looks extremely promissing for
the Standard Model indeed doing very
well in the game provided we include
``dividing out the center'' factor
$(\# center)^{2/d}$. The only two
genuine simple Lie algebras in the
Standard Model then cme out with
respectively gold and silver medals,
$SU(2)$ with gold, $SU(3)$ with silver.

\item{{\bf The problem of $U(1)$}}
With the $U(1)$ there are several
problems, which we must discuss:
\begin{itemize}
\item{1.} Since the adjoint
representation should be considered
either as non existing or as trivial
we must consider the quadratic Casimir
for the the Abelian $U(1)$ as either
$C_A(U(1)) =0$ or at best for our hopes
for favouring the Standard Model ill-
defined.

Actually there is a possibility for
making some sense of the ratio $C_A/C_F$
if we could somehow arbitrarily select
one representation of $U(1)$ given
by some ``charge'' $q_A$ to be considered
formally the ``adjoint'' $A$ and then
another one with another ``charge'' $q_F$
to be the F-representation. Then one
would naturally say that the Casimir
is the square of the ``charge'' so
that $C_A = q_A²$ and $C_F = q_F²$.
In this case of course our competition
quantity $\frac{1}{c_F} = \frac{C_A}{C_F}
= \frac{q_A²}{q_F²}$. But what shall
be considered $A$ and what $F$ ?
\item{2.} The idea that one could
``divide out the center'' of one of the
genuine simple Lie groups such as
$SU(2)$ or $SU(3)$ were meant to
mean that after having divided it out
we got instead the groups $SU(2)/Z_24$
and $SU(3)/Z_3$ instead. But then we
should only be allowed to use as $F$
the representations that are
representations of these {\em groups}.
But then the representations $F$ which
we used in the construction of our
$1/c_F$'s above are {\em not allowed}.
That in turn would mean that we would
have instead of the $F$ we used in the
cases mentioned
and actually typically to use rather
the adjoint representation itself, so
as to get for the competing quantity
$1/c_F$ now replaced by $1/c_A$ =
$C_A/C_A=1$. If we do that we loose
a factor bigger than 2 for the algebras
in the strong field. That is not
compensated by the extra factor
$(\# center)^{2/d}$ and if this
extra factor is only achievable
by paying the price that only the
adjoint representation get allowed to
be used as $F$, then it is better for
winning the game to give up the extra
factor.

\item{3.} If the total group - the
cross product say of several simple
factors - has a $U(1)$-factor in it,
one can divide out a subgroup of the
center that could be e.g. $Z_2$, if
we have $SU(2)$ and $Z_3$ if we have
$SU(3)$ in such a way that this
divided out subgroup of the center
is not subgroup neither of the genuine
simple Lie group nor of the $U(1)$
seperately. If one divides say a
$Z_3$ or a $Z_2$ out in such a way,
then it does not prevent that there
can be a representation which with
respect to $SU(2)$ or $SU(3)$ corresponds
to the $F$ we used in our above
calculation and which mannaged to make
these simple algebras win the game.

In this way we can claim that we have
a way - by means of using a $U(1)$ -
to both get the favourable $F$
representaion used to let our favourites
win, and at the {\em same time}
get ``division out of the center''
take place.

This situation seems so favorable
and really needed to get win
for a simple Lie algebra by the help of
the extra factor, that unless it
requires a very high price in form
of some loss in the final score,
it seems to be very needed to include
a $U(1)$-factor in the total group.

So here we have essentially argued
that unless the rules for the Abelian
$U(1)$ get adjusted in detail  to be very
unfavourable for winning then because
of the otherwise impossible combination
of the extra factor and the
representation $F$, it becomes needed
to have a $U(1)$ included in the total
group.

\end{itemize}

\end{itemize}

\label{scores}
\subsection{Standard Model Group
very promissing, crude review}
Let us here argue how one with very
little (extra) assumptions about the
averaging, when having a team of Lie
algebras, is to be taken, can argue
for the Standard Model group being the
winner among teams of Lie algebras:

We must of course have some rule
for making a score for a group
that is not simple fro the score
numbers for those simple invariant
subgroups of the group. One can imagine
several weightings such as e.g.
weighting the individual simple group
scores by the dimensions of the simple
groups. But of course our derivation
that the Standard Model wins the game
would be most convincing if it could
be done with so mild assumptions as
posible concerning these rules of
combining. Otherwise we could be
acused for having adjusted the rule
of weighting so as to favor the Standard
Model (if we do not succeed in arguing
that it does not matter much what rule we
use, then of course we shall assume
some rule that favours the Standard
Model, so as to see if it is at least
possible to make the Standard Model
win in such a way.)

If we cannot get the ``extra factor
from dividing out center(subgroup)''
$(\# center)^{2/d}$, the largest
achievable score for any simple
Lie algebra and therefore also for
any (sensible average, which of
course ca never be bigger than
the quantities from which it is
averaged) average over a ``team''
(a non-simple group)  becomes the
8/3 which is the biggest achievable
value for $1/c_F$, being reached for
the simple algebra $SU(2)$. To reach
a score higher than these $8/3 = 2.6667$
we have to obtain the extra factor
$(\# center)^{2/d}$, but for that we
{\em need} to have a $U(1)$. So we
suggestively should have a $U(1)$
combined with an $SU(2)$ and then have
a center $Z_2$ divided out in a way
that is not a subgroup of neither
$SU(2)$ nor $U(1)$. But that means
we have now suggestively reached
$U(2)$. It could now seemingly
be possible that what would win
would be just one or a cross product
of several $U(2)$'s. We can namely
with a sensible averaging not get
a different score for a group
and this group crossed with itself
a number of times. But now we already
argued that we needed the $U(1)$. So
we ask, could we not apply this same
$U(1)$ several times instead of just
to help $SU(2)$ to get a high score?
Actually we may use it again, but we
cannot use it to help another algebra,
which has again a $Z_2$ center to get
divied out. The problem is that if we
attempt that, we shall miss the allowance
to use the representation $F$ we used
for both the $SU(2)$ and the other
algebra, that also has the center being
$Z_2$. That woould mean that the
score for either $SU(2)$ or the other
Lie algebra would miss more than a
factor 2 in the score. If, however,
we can divide out a center which is
a $Z_n$ with an odd $n$ so that it has
no common factor with the 2 in $Z_2$,
there will be no such problem. So
e.g. a $Z_3$ wuld be o.k.. Such an
extension with a Lie algebra that
had a group from which we could divide
out e.g. $Z_3$ could be added without
the need for any further $U(1)$.
From what we already saw about the
individual scores for the simple Lie
algebras or rather groups the silver
medal winner were already $SU(3)$.
So now we must ask, if it so to speak
would pay in terms of getting the best
score average for the full group
if we to our first suggestion $U(2)$
add/extend with the $SU(3)$. Because
of the ambiguity comming from that we
do not clearly have settled how to count
the $U(1)$ we do not know, if the addition
of the proposed $SU(3)$ will pay.
It is namely so: If the averaging
of $SU(2)$ with the $U(1)$ has brought
this average from the $8/3 * 2^{2/3} =
4.233069472$
 below the score-value
of $SU(3)$ being $9/4 *3^{1/4} =
2.961166529$, then it will pay
to include the $SU(3)$.
That might happen, but we must admit
that it dependence on the
exact averaging rule, as well as on
what one puts the score for $U(1)$
in itself. So honestly we only got to,
that it is possible to imagine an
averaging proceedure, that would make the
Standard Model win!

\subsection{Extraction of the $1/c_F$}
\label{extraction}
In \cite{Rittenberg} we find for the
quadratic Casimir $C_A$
\begin{equation}
C_A= \eta g,
\label{CA}
\end{equation}
where $g$ is the dual Coxeter number,
while $\eta$ is a notation-dependent
normalization constant, which is defined
via the formula
\begin{equation}
C_R = \frac{\eta}{2}\sum_{i=1}^r \sum_{j=1}^r
(a_i +2)G_{ij}a_j
\end{equation}
for the quadratic Casimir in the
representation $R$. Here again the
quantities $a_i$ for $i= 1, 2, ...,r$
are the Dynkin labels for this
representation $R$. Finally the $r$
is the rank of the group, and $G_{ij}$
is the inverse of the Cartan matrix.

Using still \cite{Rittenberg} the ``second
index'' $I_2(V)$ for the ``vector''
representation $V$  given as
\begin{eqnarray}
I_2(V) & = & \frac{\eta}{2}
\hbox{ for SU(N) and Sp(N),}\\
I_2(V) & = & \eta \hbox{ for SO(N).}
\end{eqnarray}
and the relation
\begin{equation}
I_2(R) = \frac{N_R}{N_A} C_R,
\end{equation}
where $N_R$ is the dimension of the
general representation $R$ while
$N_A$ is that of the adjoint representation $A$, we get
\begin{eqnarray}
\frac{1}{c_F} & = & \frac{C_A}{C_F}
=\frac{C_A}{C_V}  =
= \frac{2N_Vg}{N_A} \hbox{ for SU(N)
and Sp(N),}\\
\frac{1}{c_F} &=& \frac{C_A}{C_F} =
\frac{C_A}{C_V} =\frac{N_V g}{N_A}
\hbox{ for SO(N).}
\end{eqnarray}
(We here took it that the ``smallest''
representation $F$ were indeed the
``vector'' representation $V$, which
is not always the case)
Herein we shall then insert
the dual Coxeter numbers $g$, which are
\begin{eqnarray}
g_{A_r} & = & r + 1 = N \hbox{ for }
A_r = SU(N)\hbox{ where } r=N-1,\\
g_{B_r} & = & 2r -1 = N-2 \hbox{ for }
B_r =SO(N) \hbox{ for $N$ odd and }
r =\frac{N-1}{2},\\
g_{C_r} & = & r +1 \hbox{ for symplectic
groups $C_r$}\\
g_{D_r} &=& 2r -2= N-2 \hbox{ for $N$ even
and } D_r = SO(N) \hbox{ where }
r = N/2,\\
g_{G_{2}} &=& 4 \hbox{ for} G_2,\\
g_{F_4}& =& 9 \hbox{ for } F_4,\\
g_{E_6} &=& 12 \hbox{ for } E_6,\\
g_{E_7} &=& 18 \hbox{ for } E_7,\\
g_{E_8} &=& 30 \hbox{ for } E_8.
 \end{eqnarray}
and then we obtain e.g.
\begin{eqnarray}
\hbox{For $A_r = SU(r+1)$ :}
\frac{C_A}{C_V} &=& \frac{2g_{A_r}N_V}{N_A}=
\frac{2(r+1)(r+1)}{(r+1)² -1} =
\frac{2}{1-1/(r+1)²} =
\frac{2}{1 - 1/N²}\\
\hbox{ For $B_r = SO(2r +1)$ : } \frac{C_A}
{C_V} &=& \frac{g_{B_r} N_V}{N_A}
= \frac{(2r-1)(2r+1)}{r(2r+1)} =2-1/r,\\
\hbox{ For $C_r = Sp(2r)$ : }
\frac{C_A}{C_V}&=& \frac{2g_{C_r}N_V}{N_A}
= \frac{2(r+1)2r}{r(2r+1)}
=\frac{4(r+1)}{2
r+1} = \frac{2N+4}{N+1}\\
\hbox{ For $D_r = SO(2r)$ : }
\frac{C_A}{C_V}&=& \frac{g_{D_r}N_V}{N_A}
= \frac{(2r+1)2r}{r(2r-1)} =
\frac{4r+2}{2r-1} = \frac{2N+2}{N-1}
\end{eqnarray}

But now we must admit that those ``vector''
representations $V$ which we here used
are not in all cases the smallest neither
as concerns the quadratic Casimir nor
w.r.t. dimensions. This is the case
for the relatively low rank $r$ $SO(N)$
groups. They have namely spinor
representations. In fact we have for an
even $N$ that  $SO(N)=SO(2r) = D_r$ have a
spinor representation - we shall have
the chiral irreducible representation -
of dimension $2^{r-1} = 2^{N/2 -1}$;
for odd-$N$ we have $SO(N) = SO(2r+1)
= B_r $ a spinor representation of
dimension $2^{r}=2^{(N-1)/2}$.

We may read this problem off in the
list of what \cite{Rittenberg} propose as
``reference representations''. Here the
odd-N SO(N) algebras $B_1$, $B_2$,
$B_3$, and $B_4$ are proposed represented
as reference representations by their
spinor representations, while it is
for $r \ge 5$ the $B_r$ have as their
reference representations the vector representations $V$.
Similarly it is proposed
to use as ``reference representations''
for the even SO(N)-algebras  in the
case $D_3$ meaning $SO(6)$, while
for $r\ge 4$ we use the ``vector''
representation.

\section{How to Combine Scores to Scores for Non-Simple
Groups ?}\label{averaging}
When we combine simple gauge goups
into semisimple group we have to postulate
some rule for combining and in some way
averaging our quantities for the various
simple groups. We might think of more
complicated rules but in the light of the
``theory behind'' the favoring of the
groups

\subsection{The $U(1)$ problem,
what to
take for its $C_A/C_F$ }
Since the Lie algebra of $U(1)$ has a
trivial adjoint representation it has
really no meaning to talk about $C_A$
for $U(1)$, or we might say it is zero,
but a zero is not so usefull for our
normalization.

We could propose instead to replace the
adjoint representation by the
``unit charge'' representation of the
$U(1)$ and use that as a normalization
representation. Now we should ask as we
did for the other groups: can we find
a ``smaller'' representation ? That should
now be one with a smaller charge, but
such a smaller charge would only be
allowed if we used a bigger version of
the $U(1)$ circle. So keeping the group
unchanged there are no smaller charges
allowed. Looked upon this way we can say
that the $U(1)$ is analogous to the
$E_8$ algebra for which there is no
smaller represetation than the adjoint one.
Therefore we get for $E_8$ that
$C_A/C_F =1$. Therefore we should by
analogy also take $1$ for the abelian
group $U(1)$. Then it may not matter
so much whether we really have and use
an adjoint representation.

\subsection{Volume product
weighting, a proposal}
One way to combine into some average the
scores of the different simple groups
going into
not simple Lie group is suggested by
having in
mind that
\begin{itemize}
\item{a:} We thought of the chance of
getting symmetry
by accident crudely being a good symmetry
for some a priori
``random'' action suggesting that it is
the volume of possible set of
field configurations  in which the
group transformation brings
a state around by transformation under
the group that counts. (This
volume should be minimalized to make the
chance for having the accidental
symmetry by accident maximized.)

\item{b:} We should attempt to count in
such a way that just putting
some repetition of the group as a cross
product should not change the chances;
rather it should be the type and
structure of the group occuring that we
should get information about.

\item{c:} When we have say a croz product
of groups the image
in the configuration space should also
have the character of being a product,
so that the volume of the combined group
representation would become a product
of the volumes of the components.

\item{d:} The quantity, which we
used $\frac{C_A}{C_F} *(\# center)^{2/d}$
were - by the accident of our notation -
as going inversely as the
$2/d$th power of the volume in the
configuration space relative to
some more crudely chosen group volume.
(indeed we selected this group
volume by means of the commutation rules
so as make it given by the quadratic
Casimir for the adjoint representation.)

\end{itemize}

The way suggested by this thinking is
that we should use logarithms
of our numbers used for scores and
weight them by the {\em dimensions}
of the groups. That is to say we propose
the quantity:
\begin{equation}
T =\frac{ \sum_S d_S\ln(
``\frac{C_A}{C_F}''|_S*
(\# center)^{2/d}|_S )}{
\sum_S d_S } =
\frac{1}{\sum_S d_S}\ln{ \prod_S
``\frac{C_A}{C_F}''|_S^{d_S}
(\#center_S)^2}.  \label{av}
\end{equation}

Here $S$ symbolizese the various simple
Lie algebras going into the
non-simple group, we consider. So e.g. in
the case of the Standard
Model Group $S(U(2)\times U(3))$ this $S$
runs over the
three Lie algebras,$ S = U(1), SU(2),
\hbox{ and } SU(3)$.

Having already found above the scores
for $U(1)$ being in our way counted as
$1$ meaning a $0$ when we
take the logarithm (this were somewhat not quite clean, but the
most reasonable), for $SU(2) $ the
seemingly everyone beating
$\frac{8}{3}* 2^{2/3}= 4.233069472$, and
for the $SU(3)$ score
$\frac{9}{4}*3^{2/8} =2.961166529$, we may
as an example evaluate
the by Nature beloved Standard Model
Group:
\begin{eqnarray}
T_{SMG=S(U(2)\times U(3))}& =&
\frac{1*0 + 3*\ln{ 4.233069472} +
8 *\ln{2.96116652}}{1+3+8}\\
&=& 0+0.360731843 +0.723722192= 1.084454035.
\end{eqnarray}
This means that the averaged score for
the Standard Model group
should be counted as having this averaged
quantity as its logarithm, so that
it becomes itself:
\begin{equation}
\exp{T_{SMG=S(U(2)\times U(3))}} = 2.95782451.
\end{equation}

This score by the Standard Model shall be
compared to other
obviously competing candidates such as
$U(2)$. We should remember
that without the company of the $U(1)$
the $SU(2)$ is not allowed
to gain its $2^{2/3}$-factor, so without
this $U(1)$ it would not even  have
$8/3$ = $2.666666667$(because we could
not use the representation $F$ being
the spin 1/2) and could not compete. With
the inclusion
of $SU(2)$ having to carry along the
$U(1)$ - with only its 1 score -
we get this 4.233069472 formally for
$SU(2)$ cut down to its
$3/4$th power, meaning $2.951151786$
for $U(2)$. It is really
a very tight game but it is the Standard
Model that wins over even the
$U(2)$ ! That it must be like that is
also signaled by that fact,
that the number for $SU(3)$ when the
center-factor is counted is
2.96116652 and brings the avrerage for
the Standard model group up.
This makes us look for if $U(3)$ could
now beat the Standard Model
Group? Well $U(3)$ would score the 8/9th
power of these 2.96116652
giving 2.624690339, which is less than
the score of the Standard Model
Group 2.95782451.

It should be  remembered, that the
application of this formula
should be done only, when there are
sufficient $U(1)$'s to make the
simple groups $S$ over which we sum get
their $F$-representations used
realized. It is really the importance of
the $SU(2)$ and the $SU(3)$
groups sharing their $U(1)$. This is only
possible because their
center $Z_N$'s have mutually prime numbers $n$, namely 2 and 3.
It is this collaboration between the two
by sharing the burden
of the $U(1)$ which they need for getting
their center-factors
$2^{2/3}$ and $3^{2/8}$ respectively, that
brings the Standard Model Group
$S(U(2) \times U(3))$ to win.
All of the three simple groups
collaborate  to win.

We leave it to the reader to check that
no other combination of
groups can beat the Standard Model Group!
Most of the competers are
soon loosing out, because it is only the
small rank
simple groups that get the high scores.

\section{Conclusion on the
Game Found so far}
\label{gameconclusion}
Let us summarize the most important of
the games discussed -
the game between ``teams'' meaning Groups
that are not necessesarily
simple, so that they appear as
combinations of simple algebras. Here the
proposal for game quantity is the w.r.t.
to dimension averaged logarithm of the
quantity originally
proposed $\frac{C_A}{C_F} $ including -
if allowed  without spoiling the
representation $F$ used - a
$center$-factor $(\# center)^{2/d}$. To
bring the total averaged logarithm
$T$ for a group that is typically not
simple
to be compared to the previously
discussed numbers it may be best to
exponentiate it back by taking as
the ``team-score'' (meaning score
for groups, that are not neccessarily
simple) $\exp{T}$ for the
combined group in question.

For this dimension averaged quantity we
found that the Standard Model Group
$S(U(2)\times U(3))$ (as suggested from
the representations of the
quarks and leptons found in nature) is
the maximal score of
\begin{equation}
\exp{T_{SMG=S(U(2)\times U(3))}} =  2.95782451.
\end{equation}

This $SMG = S(U(2)\times U(3))$ is
extremely tightly followed by
$U(2)$ which got
\begin{equation}
\exp{T_{U(2)}} = 2.951151786.
\end{equation}

But it were the Standard Model, that won!

If the reader would accept that the rules
of the game were chosen in
a reasonable simple way, one would say,
that it is very remarkable, that we have
been able to present a game giving just
the Standard Model Group the best
score! It should be expected there were
a reason, that should be found,
to explain, why precisely this group with
the highest score in our sense
should be the realized one. So this
finding should possibly bring us
to get to an understanding of the
question: Why the standard model group?

\section{Our Early Model with
MPP and Numerically
Maximizing Plaquette action}
\label{MPProute}
We came into the ideas of the present
article by in a lattice gauge theory
speculating
about some reason for that the energy
per plaquette normalized in some way
should be minimzed. Of course such
an energy of a plaquette contribution
to the energy depends on the
couplingsconstants, the finestructure
constants, and so we would have to combine
such a looking for minimal energy
(or a minimal action),with some assumption
about what the coupling constants would
be with different possibilities for the
guage group. As such a machinery to
provide the gauge couplings we then had
in mind to assume the idea of multiple
point principle MPP, which means in a
lattice gauge theory that the gauge
coupling parameters shall be adjusted so
as to get the lattice theory go to a
``multiple point'', i.e. a point in
coupling constant space where several
phases meet.

We shall not too deeply into the
calculations needed in the present
article. What we have to do is to
use the constraints on the coupling
constants imposed by the requirement
of the several phases just meet, that
is to say the couplings are in this sense
``critical''. In principle we can include
several possibly only as lattice artifact
relevant parameters among the here
mentioned ``coupling constants''.
Using such contraits which in principle
are constraints which we can calculate
we should be able to have estimates of
coupling constants even for groups, which
are not realized, but only thought of as
possibilities. In this way we become able
to estimate questions such as what would
the energy or action (whatever we ask for)
per plaquette in the lattice theory
be, if the group were say $G$. Thus we
can in such a scheme ask for maximizing
e.g. say the action of the plaquettes.

Now the question is, if we can make the
details of the here proposed scheme
so that we get th groups classified
much the same way as we have in the
present article proposed partly by
phenomenological guessing. Indeed it seems
that the scheme with use of MPP to
restrict the coupling constants a then
maximizing the plaquette action normalized
in an appropriate way with the square
of the dimension involved.

\subsection{On our Finestructure
Constant Fitting in New Light}
One possibly great feature of using the
sceme with MPP and maximaizing plaquette
action is that it together with the
selection of the group also provide the
coupling constants, so that we in addition
to the prediction of the gauge group
as we did in the present article get
a {\em related} prediction about the
fine structure constants. This can
hopefully soon bring us to present a
fit to the latter in such a {\em combined}
scheme. That might open up for
making interesting phenomenology on the
details of the model-type proposed by
fitting both the gauge group and the
fine structure constants.
\section{Further Speculations for a
Reason for the Selection}
\subsection{What is Good for
Prevention
of Spontaneous Breaking
of Gauge Symmetry}
We should imagine a gauge glass or just a
glassy structure in the sense
 that the action is given with terms
which vary from point or lattice cell
to point in a quenched random way. This
is what we mean here in the
abstract sense by ``glass'' that the
theory or its action involves a lot
of quenched random - meaning fixed
randomly before you integrate to make
the partition function or the Feynman
path integral - variables, so that in a
way one could think of it as, the
theory itself being random. It is
even random in a non-translational
invariant way in as far as it varies from
point to point or from little lattce
neighborhood to the next little lattice
neighborhood.

The main point, we now want to point out
is that if we let the quenched
random theory not a priori obey gauge
symmetry and gauge symmetry
has to come out the way suggested in
\cite{Foerster} the gauge theory
that we might formally think about is
also in the danger of being
broken - spontaneously - by the ground
state not having the the plaquette
variables driven to a center element - as
is required for the invariance
under a global guage transformation of the vacuum - but to some non-central
element. Honestly speaking: in the
quenched random model it will
almost certainly happen that here and
there in space(time) will
be plaquette variables, which actually
will lead to the minimum energy
density, say by standing at some
non-central element. If it stands at
a central element, it is not so serious,
since we can essentially just
think of all the elements being displaced
by a right translation and
that after such a transformation the
central element at the bottom of the
energy were transformed into the unit
element so that we can really think
of it as if it had the bottom at the unit element. But for a noncentral
element being at first at the bottom we cannot transform it to the
unit element without changing the system physically. So if truly
a non-central value occurs for the vacuum field it means indeed that
the global part of the gauge group in question has broken spontaneously.
\section{Conclusion and Outlook}
\label{conclusion}
We have in the present contribution put up an attempt to by combined
looking at some physical ideas behind and on the goal of making
the Standard Model group win produce some function defined
for compact Lie \underline{groups} with the property that it singles out
just the Standard Model group $S(U(2)\times U(3)$ as being {\em the}
Lie group for which this function has its biggest value. Indeed we
mannaged - in an almost satisfactory way - to construct such function
in a reasonable simple way. The proceedure for evaluating our proposed
function is like this:
\begin{itemize}
\item{A)} For each of the non-abelian simple Lie algebras of the Lie algebra
we construct the quantity
\begin{equation}
\frac{C_A}{C_F}* (\# center)^{2/d}, \label{quantity}
\end{equation}
where $C_A$ is the quadratic Casimir for the adjoint representation $A$
of the simple non-abelian group in question. The quadratic Casimr $C_F$
is for a ``smaller'' represetation if possible and this representation $F$
shall be chosen at the end with the purpose of making the quantity
(\ref{quantity}) as large as possible. Typically the representation $F$
will be the ``fundamental'' representation.The quantity $\# center$ is
the number of elements in the center of the covering group for the Lie algebra
in question, and $d$ is its dimension.
\item{B)}Next average the logarithm of this quantity over all the simple
non-abelian Lie weighting with the dimensions $d$ of the Lie algebras
including  the abelian components in the total Lie algebra for the
whole group counted to give 0 in logarithm (as if the quantity
(\ref{quantity}) were $1$ for $U(1)$). This average is presented in
equation (\ref{av}).
\item{C)} There is, however, an important restriction forbidding, that
unless there are enough $U(1)$'s included in the group and they have
got to a sufficient degree some (discrete) center-subgroup divided out
of the cross product of the covering groups and and the $U(1)$'s in
a way connecting the groups to be no longer just a cross product
of seperated groups, we cannot use the formula above. This restriction
shall be understood to mean:

Firstly: Under the division out of the (non-trivial) subgroup of the center
of the cross product of the abelian and non-abelian simple groups
we must not identify the center elements of any of the simple groups
so that we obtain a factor group, which no longer has the representation
$F$ for that simple groups as a representation without allowing phase
ambiguities. (This will typically mean that there must be no element
in the invariant discrete subgroup divided out which is trivial w.r.t.
to the $U(1)$'s, because that would typically lead to that the representation
$F$ would not be a representation of the \underline{group})

Secondly: In order to obtain the factor $(\# center)^{2/d}$ for one of the
simple Lie algebras - averaged betwen - it is required that the discrete
subgroup divided out has indeed a factorgroup in correspondance to the
center of the simple group in question. (This requirement implies that
the divided out discrete subgroup of the center of the product
of the covering groups and the $U(1)$'s should (at least) has as
many elements as the product of the numbers $\# center$ for all
the simple groups for which the factor $(\# center)^{2/d}$ in
(\ref{quantity} ) is to be used.).
\item{D)} The quantity - the score so to speak - which should be
largest possible for the Lie group to be realized in Nature should
under the restrictions in C) be the average constructed under B).
      \end{itemize}

The really remarkable fact of the present article is that
{\em The Standard Model Group as phenomenologically defined partly
under use of its physically realized representations of quarks and leptons
and the Higgs turns out to be precisely that (compact) Lie group which
gives the biggest value for the average constructed under B) with the
restrictions C) imposed!}

In fact one gets for the exponential of the average over the logarithms
as told in B) the number  {\bf  2.95782451}.

The Standard Model Group is, however, remarkably closely followed by the
Lie gorup $U(2)$ for which the exponetiated average becomes:  2.951151786.
They only deviate on the fourth significant ciffer, and difference is only
of the order of 0.007 compared to almost 3.

In our opinion the proceedure for constructing the function of the
compact Lie groups, the score in the game so to speak, is so simple
that one would say it is pretty remarkable that it should give just the
Standard Model Group, which is realized in Nature at least for the
energetically accessible physics in practice, to have the biggest average.
after all there are many groups which nature could have chosen, if one
did not impose the phenomenological or other restrictions. Of course the
Standard Model Group is the only fitting if we do not include parts
of the group, which are not at all seen experimentally at present.
But that just means that the Standard Model Group is - we could say - measured
to be the true model. In the present paper we search for some theoretical
assumption as simple as possible, that could single out and point to
justthis special group $S(U(2)\times U(3))$, which is the by the
representations of quarks and leptons {\em group} with the Standard Model
Lie algebra, and we found the principle of maximizing the quatity
$\exp{T}$ where $T$ is the average described! It singles out the
{\em right} group for nature!

\subsection{Taking serious that it is not an accident}

The point of such an excersie as the present one is of curse to get
some hints as to what is the reason Nature has just chosen the Stadard Model
Group and not some other group among the after all pretty many groups she
could have chosen between.

It were above suggested that the quantity in which the Standard Model
Group is excellent is that compared to a normalization given by
the quadratic Casimr $C_A$ for the adjoint representation $A$ the group
has (a) very small representation(s) in terms of some quadratic Casimir
$C_F$ for a representation, which we above have thought upon as a
representation related to the fields in the e.g. lattice guage theory
model working in Nature. The thing that seems to be important is that
compared to some ``natural'' distance measure on the group
(related to the quadratic Casimir for the adjoint representation $C_A$)
the way it is possible to make it move the fields in some appropriate
representation $F$ is very slow. That is to say you may move the group
element a lot but the fields only tinily for the groups having high scores in
our game. Such a property of it being easy to push the feilds only little
around for the group element moving much without making the transformation
completely zero ( i.e. still usinga true representation $F$) seems to be
what our result points to as the important principle used to select just the
model, which nature has chosen.

We suggested that such a selection were likely to be the result, if
the gauge group had in reality appeared by first getting an approximate
gauge group by accident. Then the gauge symmetry should for practical
purposes have become exact due to quantum fluctuations. But the important
point to extract is that the choice of the Standard Model group suggests
that the group that can be represented, on fields say, being most tinily
moved around under a by some adjoint representation related normalization of
distances in the group  is the group most beloved by nature to be realized.
\section{Acknowledgement}

\begin{figure}
\centerline{\includegraphics[height=10cm,scale=0.5]{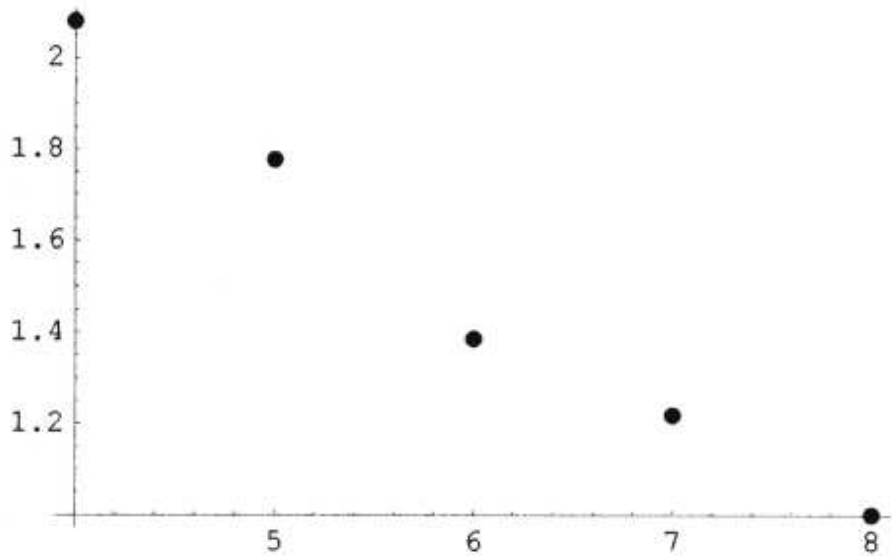}}
\caption{\label{fig} The ratio $C_A/C_F$ for $E$ groups plotted as a function of rank. Here we have used that
$E_5=SO(10)$ and $E_4=SU(5)$. }
\end{figure}

One of us (H.B.N.) thanks for being allowed to have room and
stay as emeritus at the Niels Bohr Institue at Copenhagen
University, since 1st of September, when the writing of this
proceeding took place.

\section{Appendix: A Pedagogical
Calculation Procedure for our Purpose} Having in mind that one
of the main ideas for why our proposed quantity $C_A/C_F$ -
i.e. the ratio of the quadratic Casimir for the adjoint
representation $C_A$ divided by some representation $F$ having
the smallest quadratic Casimir $C_F$ among faithfull
representations  - is suggested to be that this quantity
$C_A/C_F$, the bigger it, is favours the chanse that the group
in question should be the one realized in nature
  because a big $C_A/C_F$ means that
varying the potential gauge an amount
measured by the Killing form normalized
by the adjoint representation Casimir
being ut to say 1 makes the variation
in the link variables supposedly in the
representation $F$ minimal, we shall
here present as a couple of examples
a practical calculation of our quantity,
almost making clear its physical
significance for our purpose.

Remember that we in our speculative
arguments for which group would most
easily become a gauge group by
accidentally being so near to being it
that a quantum fluctuation effect might
set in and make it practically an exact
gauge symmetry we used the normalization
of the Killing form to make the quadratic
Casimir for the adjoint representation
say 1 so that we thereby got a physically
meaningfull distance concept on the group
manifold. Under use of this distance
concept we then asked how the field theory
variables  say the link variables in some
lattice gauge theory formulation will
vary for a given - unit- variation
of the gaug group element. If one in the
link variable space use the distance
concept derived from the trace of the
square of difference of the  couple of
represntation matrices corresponding to
the link variable, the ratio of the
infinitesimal distance in the group
manifold relative to the corresponding
distance in the link variable will be
given by the square root of the ratio
$C_A/C_F$, where $F$ is the represntation
used to represent the link variable.
The main idea were that the chance to
find an approximate symmetry under the
transformation of the various link
variables as transformed under the
gauge transformation will be bigger the
bigger the ratio $C_A/C_F$ and so the
group ``to be realized by accident''
with best chnace is expected to be the
group with largest $C_A/C_F$- value.

In this appendix we shall present a way
to calculate this ratio $C_A/C_F$ by
using just the Cartan algebra
representaions in a way that allows us
to calculte simply the ratio of the
average of the square of the root vector
length compared to the corresponding
average of the weight vectors for the
representation $F$.

Let us provide a couple of examples:

\begin{itemize}
\item{$A_1=SU(2)$:}

In this case the smallest represntation
- w.r.t. say the quadratic Casimir -
is $F=\underline{2}$. As is wellknown
the root system for the $F=\underline{2}$
consists of two weights both of half
length of the roots. Thus we find if we
say roots have length $\sqrt{2}$ - as
is usual -
\begin{itemize}
\item{ Adjoint :} Average of the squared
roots $\frac{2+2+0}{3} = \frac{4}{3}$.
\item{$F$:} Averrage of the squared
weights $\frac{1/2 +1/2}{2} = \frac{1}{2}$. \end{itemize}
 So we obtain by taking the ratio of
these averages:
\begin{equation}
\frac{C_A}{C_F}|_{A_1} = \frac{4/3}{1/2} =
\frac{8}{3}
\end{equation}

\item{$G_2$ : } The root system for
the exceptional Lie group $G_2$ consists
of two regular hexagons with centers
in the zero point, the one rotated by
$30^0$ w.r.t. the other one and one
$1/\sqrt{3}$ times the other one.
The Lie algebra of $G_2$ has an
$SU(3) = A_2$ subgroup corresponding to
the roots of the bigger one of the two
hexagones. The ``smallest'' representation
$F$ for the $G_2$ is sevendimensional
and consists w.r.t. the $SU(3)$ subgroup
of a triplet an antitriplet and a singlet.
This means that the weight system for this
representation $F$ consists of the six
roots in the smaller of the two hexagons
and in addition a weight at zero.
Then we have for the average of the
squares of the distances from zero for
the weights
\begin{itemize}
\item{Adjoint representation:}
$average = \frac{6*2 + 6*2/3}{14} =\frac
{8}{7}$.

\item{$F$ :}
$average
= \frac{6*2/3 +0}{7} = \frac{4}{7}$
\end{itemize}
We thus find that
\begin{equation}
\frac{C_A}{C_F}|_{G_2} = \frac{8/7}{4/7} = 2
\end{equation}
\item{$B_2=SO(5) = C_2 = Sp(4)$ :}
The root system for these isomorphic
Lie algebras consists of the corners
and the midpoints of the sides of a
square (with side 2 say). There are thus
8 roots. The ``smallest'' representation
$F$ is a four dimensional one with the
roots with the weights sitting in the four
centers for the four squares with side 1
into which the coordinate axes divide the
mentioned square of side 2. Then we get
for the averages of the squaes of the
distances from the zero in the root and
weight systems:
\begin{itemize}
\item{Adjoint :}
$average = \frac{4*1 + 4*2+ 2*0}{10}
= \frac{6}{5}$
\item{$F$ :}
$average = \frac{4 *1/2}{4} = \frac{1}{2}$
\end{itemize}
So our competition number becomes
\begin{equation}
\frac{C_A}{C_F}|_{B_2} = \frac{6/5}{1/2} =
\frac{12}{5}
\end{equation}
\item{$A_2 = SU(3)$ : }
For $SU(3)$ the root system is a
regular hexagon around zero, and we take
the length of the roots as usual to be
$\sqrt{2} $. The ``smallest''
representation $F$ is  the quark or
we can equally well take the antiquark
representation $\underline{3}$. The weight
system for say the quark representation
forms  is a triangle centered around
zero and having the side length $\sqrt{2}$
like we took the roots to have. Thereby
the  distances of the weights from
zero become $\sqrt{2}{3}$. So the
averages of the squares of the distances
from zero becomes:
\begin{itemize}
\item{Adjoint:}
$average = \frac{6*2 +2*0}{8} =
\frac{3}{2}$
\item{$F$= ``quark''represntation :}
$average = \frac{3*2/3}{3} = \frac{2}{3}$
\end{itemize}
 So we obtain for our ratio
\begin{equation}
\frac{C_A}{C_F}|_{A_2}= \frac{3/2}{2/3}
=\frac{9}{4}
\end{equation}
\item{ $F_4$ :} The root system for
the exceptional Lie algebra $F_4$, $\Phi$,
is described as contained in $V={\bf R}^4$
and consisting of those vectors $\alpha$
with length $1$ or $\sqrt{2}$
for which the coordinates obey that
$2\alpha$ having all coordinates integer
and that so that for each $2\alpha$ these
coordinates are either all even or all
odd. There are 48 roots in this system.

These 48 roots are easily seen to
fall into one group of 16 of length 1 for
which the
coordinates are all $\pm 1/2$, one group
of 24 of length $\sqrt{2}$ having two
coordinates 0 and two $\pm 1$, and
8 roots have just one coordinate equal
to $\pm 1$ and the other coordinates
being 0.

The average square distance of these
roots together with the 4 Cartan group
basis vectors with 0 distance so to speak
becomes
$\frac{16*1 + 8*1 + 24 *2 + 4*0}{48 +4}
=\frac{72}{52}=\frac{18}{13}$.

\end{itemize}

We used here the Cartan algebra only,
but since these Cartan algebra elements
can be transformed around to go into
the non-Cartan algebra so at the end the
average ``charges'' must be the same and
thus this restriction would not matter.

\section{Appendix 2: Calculation of $C_A/C_F$.}\label{a2}
In order to calculate the ratios of
quadratic Casimirs we shall here rewrite
a list of the adjoint representaions
for the Lia algebras:
\begin{eqnarray}
A_n \hbox{ : } &(1,0,0, ...,0,0,1)&
n(n+2)\\
B_n \hbox{ : } &(0,1,0,...0,0,0) &
n(2n+1)\\
C_n \hbox{ : } &(2,0,0,...,0,0,0)&
n(2n+1)\\
D_n \hbox{ : } &(0,1,0,...,0,0,0)&
n(2n-1)\\
G_2 \hbox{ : } & (1,0) & 14\\
F_4 \hbox{ : } & (1,0,0,0) & 52\\
E_6 \hbox{ : } & (0,0,0,0,0,1)& 78 \\
E_7 \hbox{ : } & (1,0,0,0,0,0,0)& 133\\
E_8 \hbox{ : } & (0,0,0,0,0,0,1,0) & 248
\end{eqnarray}

Here the orders of the Dynkin labels
correspond to enumerating the being
successive in the chain except for the
$E$-algebras for which the largest number
though is assigned to the node which
is both an end node and attached to the
node having three neighbours. In the cases
of $B_n$ and $C_n$ it is the $n$th
node that is respectively the short
and the long simple roots. In cases
$F_4$ and $G_2$ the short roots are
numbered with the largest numbers.

In the same notation we also copy in
what we can call Simple irreducible
representations of the simple Lie
algebras:
\begin{eqnarray}
\hbox{For } A_n &(1,0,...,0,0)&
\quad dim=n+1\\
\hbox{or}       &(0,0,...,0,1)&
\quad dim=\overline {n+1}\\
\hbox{For } B_n & (1,0,...,0)&\quad
dim=2n+1\\
\hbox{and} &(0,0,...,0,1)& \quad 2^n\\
\hbox{For } C_n & (1,0,...,0,0) & \quad
dim=2n\\
\hbox{For } D_n & (1,0,...,0,0) & \quad
dim = 2n\\
\hbox{and} & (0,0,...,0,1) &\quad dim=
2^{n-1}\\
\hbox{or} & (0,0,...,0,1,0) &\quad
dim =2^{n-1}\\
\hbox{For } G_2 & (0,1) &\quad dim = 7\\
\hbox{For } F_4 & (0,0,0,1) &\quad dim
= 26\\
\hbox{For } E_6 & (1,0,0,0,0,0) & \quad
dim = 27\\
\hbox{or} &(0,0,0,0,1,0)& \quad dim=
\overline{27}\\
\hbox{For } E_7 & (0,0,0,0,0,1,0)& \quad
dim = 56\\
\hbox{For } E_8 & (0,0,0,0,0,0,1,0) &
\quad dim = 248
\end{eqnarray}

In order to use the equation
\begin{equation}
C_R = \frac{\eta}{2}\sum_{i=1}^r \sum_{j=1}^r
(a_i +2)G_{ij}a_j
\end{equation}
for the quadratic Casimir $C_R$ of a
representation $R$ being in our cases
of interest, we must know the matrix
elements of the ``Metric tensors for
the weight spaces''$G_{ij}$
(or the inverse Cartan matrix) at the
relevant places:
For  $R$ being the adjoint
representation $A$ we have
for the $A_n$ Lie algebra
both $a_1 =1$ and $a_n=1$ while the
other Dynkinlabels $a_i=0$. For the other
algebras than the $A_n$-series, we have
only one Dynkin label different from zero,
and that is
\begin{center}
{\bf For Adjoint Representations}
\end{center}
\begin{eqnarray}
B_n \hbox{ : }& a_2(Adj\quad B_n)& = 1;\\
C_n \hbox{ : }& a_1(Adj \quad C_n) & =2;\\
D_n \hbox{ : }& a_2(Adj \quad D_n) & =1;\\
G_2 \hbox{ : }& a_1(Adj \quad G_2) & =1;\\
F_4 \hbox{ : }& a_1(Adj \quad F_4) & =1;\\
E_6 \hbox { : } & a_6(Adj \quad E_6)&
=1;\\
E_7 \hbox{ : } & a_1(Adj \quad E_7) & =1;\\
E_8 \hbox{ : } & a_7(Adj \quad E_8) & =1.
\end{eqnarray}

For the simple representations mentioned
in the list above we have correspondingly
that the only non-zero Dynkin labels
are
\begin{center}
{\bf For simple representations:}
\end{center}
\begin{eqnarray}
\hbox{For } A_n:&\quad a_1(A_n,n+1)=&1,\\
\hbox{or}   & a_n(A_n,\overline{n+1}) =&1
;\\
\hbox{For } B_n :& \quad a_1(B_n,2n+1)&
=1,\\
\hbox{and for the spinor rep.} & \quad
 a_n(B_n, 2^n) =& 1;\\
\hbox{For } C_n : & \quad a_1(C_n, 2n) =& 1
;\\
\hbox{For }D_n : & \quad a_1(D_n,2n)=& 1,\\
\hbox{and for spinors}&\quad
a_n(D_n,2^{n-1})=&1,\\
\hbox{or} & \quad a_{n-1}(D_n, 2^{n-1},*) = &
1;\\
\hbox{For } G_2:& \quad a_2(G_2, 7)=&1;\\
\hbox{For } F_4:& \quad a_4(F_4, 26) = &
1;\\
\hbox{For } E_6: \quad a_1(E_6, 27) = &
1,\\
\hbox{or} & \quad a_5(E_6, \overline{27})
=&1;\\
 \hbox{For } E_7: \quad a_6(E_7,56)=&1;\\
\hbox{For } E_8: \quad a_7(E_8, 248) =&1
\end{eqnarray}
So except for the case of $A_n$, in which
we need the $G_{1,n}= \frac{1}{n+1}$ and
$G_{n,1}=\frac{1}{n+1}$ matrix
elements also, we only need the diagonal
elements
and the sums of the elements in the
columns
of the metric tensor matrices for the
weight or the inverse Cartan matrices.
We therefore here present these diagonal
series of elements:
\begin{center}
{\bf Diagonal Elements of Weight Space
Metric}
\end{center}
\begin{eqnarray}
A_n&:& G_{1,1}=\frac{1*n}{n+1}, G_{2,2}=\frac{2(n-1)}{n+1}, ..., G_{(n-1),(n-1)}=\frac{(n-1)*1}{n+1}, G_{n,n}=\frac{n*1}{n+1};\\
B_n&:& G_{1,1} = 1, G_{2,2}=2,...,
G_{(n-1),(n-1)}=n-1, G_{n,n}=\frac{n}{4};\\
C_n&:& G_{1,1} = \frac{1}{2}, G_{2,2}=1,...
G_{(n-1),(n-1)}= \frac{n-1}{2}, G_{n,n}=
\frac{n}{2};\\
D_n & : & G_{1,1}=1,
G_{2,2}=2,..., G_{n-2,n-2}= n-2, G_{n-1,n-1}=
\frac{n}{4}, G_{n,n}=\frac{n}{4};\\
G_2&:& G_{1,1} = 2, G_{2,2}=\frac{2}{3};\\
F_4&:& G{1,1}= 2, G_{2,2}= 6, G_{3,3}= 3,
G_{4,4}= 1;\\
E_6&:& G_{1,1}=\frac{4}{3}, G_{2,2}=
\frac{10}{3}, G_{3,3}=6, G_{4,4}=
\frac{10}{3}, G_{5,5}=\frac{4}{3}, G_{6,6}=
2;\\
E_7&:& G_{1,1}= 2, G_{2,2}= 6, G_{3,3}=12,
G_{4,4}= \frac{15}{2}, G_{5,5}= 4, G_{6,6}=
\frac{3}{2}, G_{7,7}= \frac{7}{2};\\
\vspace{-1mm}E_8&:& \vspace{-1mm}G_{1,1} =4, G_{2,2}=14, G_{3,3}=30,
G_{4,4}= 20, G_{5,5}= 12, G_{6,6}=6,
G_{7,7}=2, G_{8,8}=8;
   \end{eqnarray}

In addition we needed the sums
over the columns and thus we present
these sums e.g.
\begin{equation}
Sum(A_n) = (\sum_{i=1}^{n} G_{i,1}, \sum_{i=1}
^nG_{i,2},..., \sum_{i=1}^nG_{i,n})
\end{equation}
and get the following:
\begin{eqnarray}
Sums(A_n) &=& \left(\frac{n}{2}, n-1,
\frac{3n}{2} -3, ...,n-1 ,\frac{n}{2}
\right) \\
Sums(B_n) &=&\left ( 1(n-1/2), 2(n-1),
3(n-3/2), ...
,\frac{(n+1)(n-1)}{2}= (n-1)(n-(n-1)/2),
\frac{n²}{4}\right )\\
Sums(C_n) &=& \left(\frac{n}{2},
\frac{2n-1}{2}, \frac{3n-3}{2},...,
\frac{(n+2)(n-1)}{4},\frac{(n+1)n}{4}
\right)\\
Sums(D_n) &=& \left ( n-1, 2n-3,3n-6,...,
\frac{(n-1)n}{2}, \frac{(n-1)n}{2}
\right )\\
Sums(G_2)& =& \left ( 3, \frac{5}{3} \right )
\\
Sums(F_4)& =& \left ( 8, 15, \frac{21}{2},
\frac{11}{2} \right )\\
Sums(E_6)& = & \left ( 8,15, 21, 15,8,11
\right )\\
Sums(E_7)& =& \left ( 17, 33, 48,
\frac{75}{2}, 26, \frac{27}{2},
\frac{49}{2} \right )\\
Sums(E_8) & =& \left ( 46, 91, 135, 110,
84, 57,29, 68\right )
\end{eqnarray}

Denoting the $j$th element in these
Sums by an index $j$ like e.g.
$Sums(G_2)_j$ we can then write the
expression for the typical cases above
of a ``simple'' representation where the
$a_j$ alone is different from zero:
\begin{equation}
C_F = \frac{\eta}{2}(G_{j,j} + 2Sums_j)
\end{equation}
We can thus by insetion obtain:
\begin{center}
{\bf Simple Quadratic Casimirs}
\end{center}
\begin{eqnarray}
C_F(A_n) &=& \frac{\eta}{2} (\frac{n}{n+1}
+2*\frac{n}{2}) = \frac{\eta}{2} *
\frac{n(n+2)}{n+1}\\
C_{F \; vector}(B_n) &=& \frac{\eta}{2}
(1 + 2*(n-\frac{1}{2}) = \eta*n
\hbox{for 2n+1}\\
C_{F \; spinor}(B_n)&=& \frac{\eta}{2}
(\frac{n}{4} + 2* \frac{n^2}{4} )
= \frac{\eta* (n+2n²)}{8}\\
C_F(C_n) &=& \frac{\eta}{2}(\frac{1}{2}+
2*\frac{n}{2})= \eta*\frac{(2n+1)}{4}\\
C_{F \; vector}(D_n) &=& \frac{\eta}{2}
( 1 + 2*(n-1)) = \eta*(n-1/2)\\
C_{F \; spinor}(D_n) &=& \frac{\eta}{2}
(\frac{n}{4} +2*\frac{n(n-1)}{4})
= \eta*\frac{2n^2 -n}{8}\\
C_F(G_2) &=& \frac{\eta}{2} (\frac{2}{3}+
2*\frac{5}{3}) = \eta*2\\
C_F(F_4) &=& \frac{\eta}{2} (1 + 2*
\frac{11}{2}) = \eta*6\\
C_F(E_6) &=& \frac{\eta}{2}(\frac{4}{3}+
2*8) = \eta*\frac{26}{3}\\
C_F(E_7) &=& \frac{\eta}{2}(\frac{3}{2}+
2*\frac{27}{2})= \eta*\frac{57}{4}\\
C_F(E_8) &=& \frac{\eta}{2}(2 + 2*29)
= \eta*30
\end{eqnarray}

These quadratic Casimirs can be compared
with e.g. the corresponding ones
for the adjoint representations, and then
the normalization - symbolized by the
factor $\eta$, which we thus avoid
having to choose. Thereby we obtain
the ratio which were our first proposal
in the present article for quantity
about which to hold the game.
These adjoint representation quadratic
Casimirs become:
\begin{center}
{\bf Quadratic Casimirs for Adjoint
Represntations}
\end{center}
\begin{eqnarray}
C_A(A_n) &=& \frac{\eta}{2}( 2*Sums(A_n)_1
+ 2*Sums(A_n)_n + G_{1,1}(A_N) + G_{n,n}(A_n)
+ G_{n,1}(A_n) + G_{1,n}(A_n))\\
& =&
\frac{\eta}{2}(2*\frac{n}{2}+2*\frac{n}{2}
+\frac{n}{n+1} +\frac{n}{n+1} +
\frac{1}{n+1} +\frac{1}{n+1}) =
\eta*(n+1)\\
C_A(B_n) &=&\frac{\eta}{2} ( 2*Sums(B_n)_2
+ G_{2,2})= \frac{\eta}{2} (2*2(n-1) + 2)
= \eta *(2n-1)\\
C_A(C_n) &=& \frac{\eta}{2}
(2*2*Sums(C_n)_1 + 2^2 * G_{1,1}) =
\frac{\eta}{2} (4*\frac{n}{2} + 2^² *
\frac{1}{2}) = \eta*(n+1)\\
C_A(D_n) &=& \frac{\eta}{2}(2*Sums(D_n)_2
+ G_{2,2}) =\frac{\eta}{2}(2*(2n-3) + 2)
= \eta*2(n-1)\\
C_A(G_2) &=& \frac{\eta}{2}(2*Sums(G_2)_1
+ G_{1,1} ) =\frac{\eta}{2}(2*3 + 2) = \eta
* 4\\
C_A(F_4) &=& \frac{\eta}{2} (2*Sums(F_4)_1
+ G_{1,1}) =\frac{\eta}{2}(2*8 + 2) = \eta
*9\\
C_A(E_6) &=& \frac{\eta}{2}(2*Sums(E_6)_1
+ G_{1,1}(E_6)) = \frac{\eta}{2}(2*11 +
2) =\eta* 12\\
C_A(E_7) &=& \frac{\eta}{2}
(2*Sums(E_7)_7 + G_{7,7}(E_7)) =
\frac{\eta}{2}(2*17 +
2) = \eta*18\\
C_A(E_8) &=& \frac{\eta}{2}(
2*Sums(E_8)_7
+ G_{7,7}(E_8)) = \frac{\eta}{2}( 2*29 +2)
=\eta* 30\\
       \end{eqnarray}
(These numbers are indeed $\eta*g$,
where $g$ is the dual coxeter number,
see above)
Combining these calculations of
quadratic casimirs we then finally obtain
by taking the ratios our competition
quantity $C_A/C_F$.
\begin{center}
{\bf Our Ratio of Adjoint to ``Simplest''
Quadratic Casimirs $C_A/C_F$}
\end{center}
\begin{eqnarray}
\frac{C_A}{C_F}|_{A_n} & =&
\frac{2(n+1)^2}{n(n+2)} =
\frac{2(n+1)^2}{(n+1)^2 -1} =
\frac{2}{1-\frac{1}{(n+1)^2}}\\
\frac{C_A}{C_{F \; vector}}|_{B_n}& =&
\frac{2n-1}{n}= 2 - \frac{1}{n}\\
\frac{C_A}{C_{F \; spinor}}|_{B_n}&=
&\frac{2n-1}{\frac{2n^2 +n}{8}} =
\frac{16n -8}{n(2n+1)}\\
\frac{C_A}{C_F}|_{C_n} &=&
\frac{n+1}{n/2 +1/4} =
\frac{4(n+1)}{2n+1}\\
\frac{C_A}{C_{F \; vector}}|_{D_n}&=&
\frac{2(n-1)}{n-1/2}=
\frac{4(n-1)}{2n-1}\\
\frac{C_A}{C_{F \; spinor}}|_{D_n}&=&
\frac{2(n-1)}{\frac{2n^2-n}{8}}
= \frac{16(n-1)}{n(2n-1)}\\
\frac{C_A}{C_F}|_{G_2} &=& \frac{4}{2} =2\\
\frac{C_A}{C_F}|_{F_4} &=& \frac{9}{6} =
 \frac{3}{2}\\
\frac{C_A}{C_F}|_{E_6} &=&
\frac{12}{\frac{26}{3}} = \frac{18}{13}\\
\frac{C_A}{C_F}|_{E_7}&=&
\frac{18}{\frac{57}{4}} = \frac{72}{57}
= \frac{24}{19}\\
\frac{C_A}{C_F}|_{E_8}&=& \frac{30}{30} =1
\end{eqnarray}
\section{Appendix 3: Checks and
overview
of $C_A/C_F$}
It may be comforting that one can
put the calculations in section \ref{a2},
 i.e. appendix 2, up to a few cross
checkings, such as checking that
isomorphic algebras give the same ratio
$C_A/C_F$ as of course they shall
for a notaton independent quantity:
\begin{itemize}
\item{$A_1\approx B_1 \approx C_1$}

\begin{eqnarray}
\frac{C_A}{C_F}|_{A_1}& =& \frac{8}{3},\\
\frac{C_A}{C_{F \; spinor}}|_{B_1}&=&
\frac{16*1 -8}
{1*(2*1+1)} = \frac{8}{3},\\
\frac{C_A}{C_F}|_{C_1} = \frac{4(1+1)}
{2*1+1} = \frac{8}{3}.
\end{eqnarray}

\item{$A_1 \times A_1 \approx D_2$}

\begin{eqnarray}
\frac{C_A}{C_F}|_{A_1} &=& \frac{2}{1-
\frac{1}{(1+1)^2}} = \frac{8}{3}\\
\frac{C_A}{C_{F \; spinor}}|_{D_2} &=&
\frac{16(2-1)}{2*(2*2-1)} =\frac{8}{3}
\end{eqnarray}

\item{$ B_2=SO(5) \approx C_2=Sp(4)$}

\begin{eqnarray}
\frac{C_A}{C_{F \; spinor}}|_{B_2} &=&\frac{
16*2-8}{2*(2*2+1)} = \frac{12}{5}\\
\frac{C_A}{C_F}|_{C_2} &=& \frac{4*(2+1)}{
2*2+1} = \frac{12}{5}
\end{eqnarray}

\item{$D_3 =SO(6) \approx A_3 = SU(4)$}

\begin{eqnarray}
\frac{C_A}{C_{F \; spinor}}|_{D_3} &=&\frac{
16(3-1)}{3*(2*3-1)} = \frac{32}{15}\\
\frac{C_A}{C_F}|_{A_3}= \frac{2*(3+1)^2}
{(3+1)^2 -1} = \frac{32}{15}.
\end{eqnarray}

\end{itemize}

Further we should note that for $D_4 =
SO(8)$ (w.r.t. Lie algebra) there is
symmetry between the spinor and vector
representations, which are both
8-dimensional. Thus we should have
$\frac{C_A}{C_{F \; spinor}}|_{D_4}
= \frac{C_A}{C_{F \; vector}}|_{D_4}$.
Indeed we find
\begin{eqnarray}
\frac{C_A}{C_{F \; spinor}}|_{D_4}&=&
\frac{16(4-1)}{4*(2*4 -1)} =
\frac{12}{7}\\
\frac{C_A}{C_{F \; vector}}|_{D_4}&=&
\frac{4(4-1)}{2*4 -1} = \frac{12}{7}.
\end{eqnarray}

We should also expect  approximately
the same large $N$ behavior behavior
for $SO(N)$, whether it be for even
$N$ for which we have $D_{N/2}$, or
for odd $N$ for which we have
$B_{(N-1)/2}$. Let us indeed formally
consider these two Lie algebras:
\begin{eqnarray}
\frac{C_A}{C_{F \; vector}}|_{D_{N/2}}
&=& \frac{4(n-1)}{2n-1} =\frac{4(N/2-1)}
{2N/2 -1} = \frac{2N-4}{N-1}\\
\frac{C_A}{C_{F \; vector}}|_{B_{(N-1)/2}}&=&
\frac{2n-1}{n}= \frac{(N-1) -1}{(N-1)/2}
= \frac{2N-4}{N-1}.
\end{eqnarray}
  Remarkable we get even exactly the
same formal expressions $\frac{2N-4}{N-1}$.

Similarly we may compare the spinor
representation for $F$ using ratios
$C_A/C_{F \; spinor}$ for $B_{(N-1)/2} $
and $D_{N/2}$:
\begin{eqnarray}
\frac{C_A}{C_{F \; spinor}}|_{B_{(N-1)/2}}&=&
\frac{16n-8}{n(2n+1)} =
\frac{16*(N-1)/2-8}{(N-1)/2 *(2(N-1)/2+1)}
= \frac{8N-16}{(N^2-N)/2} = \frac{16(N-2)}
{N(N-1)}\\
\frac{C_A}{C_{F\; spinor}}|_{D_{N/2}}&=&
\frac{16(n-1)}{n(2n-1)} = \frac{16(N/2-1)}
{N/2 *(N-1)} = \frac{16(N-2)}{N(N-1)}
\end{eqnarray}
So in spite of the fact that the
dimensionallity of the spinor
representations is not a smooth functon
of $N$ but rather jumps up and down
with the even or oddness of $N$, we
got formally the same formula for our
ratio for competition becomes the same
written as a function of the $N$ of
$SO(N)$.

\subsection{The speculation of the
high rank groups almost giving
same $C_A/C_F$}
We have already seen that for large rank
$r$ the infinite series of Lie algebras
have our $C_A/C_F$ going to 2. This is
not so surprising from the thinking that
as the rank goes up the root systems and
the weight system for $F$ (the
``smallest'' representation) become more
and more rich in number of roots and
weights as the rank goes up. But then
we might consider the root and weight
distributions to be more and more
statistical understandable. And if so
then we  might expect that the small
details in the Dynkin diagram deviating
from just a long chain of single line
connected nodes like in the $A_n$'s
would have less and less effect and
so the approach to a single number
common for all Lie algebras.

\end{document}